%% file: main.tex
\documentclass[letterpaper]{article}

\usepackage{prepr}

\newcommand{\acro}[0]{\textsl{FiTS}\xspace}
\newcommand{\rqtext}{How can a scenario-based approach be effectively used to represent the knowledge required for field testing a drone swarm in an environment that requires capabilities for adaptation?}
\newcommand{\mar}[1]{\texttt{<#1>}}

\title{Scenario-Based Field Testing of Drone Missions}
\author{
\PREPauthor{Michael Vierhauser}{University of Innsbruck, Department of Computer Science,\newline  Innsbruck, Austria}{Michael.Vierhauser@uibk.ac.at}
\PREPauthor{Kristof Meixner}{Institute of Information Systems Engineering, TU Wien, and CDP,\newline  Vienna, Austria}{{firstname.lastname}@tuwien.ac.at}
\PREPauthor{Stefan Biffl}{Institute of Information Systems Engineering, TU Wien, and CDP,\newline  Vienna, Austria}{{firstname.lastname}@tuwien.ac.at}
\vspace{0.6cm}
}
\setvenue{the \vspace{3pt}\textbf{\\50th Euromicro Conference on Software Engineering and Advanced Applications (SEAA)}}
\setvenueone{the \textbf{50th Euromicro Conference on Software Engineering and Advanced Applications (SEAA)}}
\setyear{2024}
\setdoi{10.1109/SEAA64295.2024.00012}

\begin{document}

\maketitle

 \input{sec_00_abstract}

%%% Sections %%%
\input{sec_01_introduction}
\input{sec_02_example}

\input{sec_03_approach}
\input{sec_05_eval}
\input{sec_06_relwork}

\input{sec_07_conclusion}

 \section*{Acknowledgments}

The financial support by the Christian Doppler Research Association, the Austrian Federal Ministry for Digital and Economic Affairs, and the National Foundation for Research, Technology and Development is gratefully acknowledged.
This work has been partially supported and funded by the Austrian Research Promotion Agency (FFG) via the Austrian Competence Center for Digital Production (CDP) under contract number 881843.

\balance
\bibliographystyle{abbrv}
\bibliography{refs}

\end{document}

%% file: sec_00_abstract.tex
\begin{abstract}

% Context. 
Testing and validating Cyber-Physical Systems (CPSs) in the aerospace domain, such as field testing of drone rescue missions, poses challenges due to volatile mission environments, such as weather conditions.
While testing processes and methodologies are well established, 
structured guidance and execution support for field tests are still weak.
This paper identifies requirements for field testing of drone missions and introduces the \textit{Field Testing Scenario Management (\acro)} approach for adaptive field testing guidance.
\acro aims to provide sufficient guidance for field testers as a foundation for efficient data collection to facilitate quality assurance and iterative improvement of field tests and CPSs.
\acro shall leverage concepts from scenario-based requirements engineering and Behavior-Driven Development to define structured and reusable test scenarios, with dedicated tasks and responsibilities for role-specific guidance.
We evaluate \acro by (i) applying it to three use cases for a search-and-rescue drone application to demonstrate feasibility and (ii) interviews with three experienced drone developers to assess its usefulness and collect further requirements.
The study results indicate \acro to be feasible and useful to facilitate 
drone field testing and data analysis.
\end{abstract}
\vspace{0.8em}

% Emerging incidents and failures in Cyber-Physical Systems (CPSs) are typically difficult to diagnose and understand as errors may appear in the software, at the hardware level, or the intersection between hardware and software, making thorough testing and validation a crucial task. 
%
% Motivation. 
% Testing and validation of CPSs is typically performed at various levels, and in addition to ''traditional'' testing methods, simulations, and ultimately real-world Field Testing are employed.
% However, 
%
% In this paper, we discuss the concepts of a novel framework \acro, providing support for the structured definition, execution, and analysis of field tests, focusing on small uncrewed aerial systems (sUAS) -- commonly referred to as drones.
% In this initial work, towards establishing the \acro framework, we focus on the first phase, the creation, management and organization of field test scenarios.

%% file: sec_01_introduction.tex
\section{Introduction}
\label{sec:introduction}

% \emph{Problem and Motivation:}
Issues and failures in Cyber-Physical Systems (CPSs), 
such as issues with drones on rescue missions, 
encompass both software-related bugs and hardware-related malfunctions~\cite{vierhauser2021hazard}. These issues pose significant risks, potentially causing serious incidents that endanger physical infrastructure and human lives.
To prevent such incidents, extensive simulations shall ensure that CPSs behave according to their specified requirements~\cite{bringmann2008model,bondi2018airsim}.
Simulations are typically employed for small uncrewed aerial systems (sUAS) -- commonly referred to as drones -- to validate system behavior in a ``safe'' environment without risking damage to buildings or the environment, or causing harm to human operators and bystanders~\cite{agrawal2023requirements,bondi2018airsim}.
However, while various CPS simulation environments exist, ranging from low-fidelity, basic simulations to high-fidelity simulations in a realistic 3D environment, there is still a gap between virtual hardware and how real systems behave in the physical world.
This simulator-to-reality gap has been recognized as a critical issue in robotics research~\cite{jakobi1995noise}, describing the mismatch of simulated and real physical-law performances caused by the inaccurate representation of the real environment in simulation~\cite{salvato2021crossing}.
Furthermore, diverse environmental factors, such as temperature, wind conditions, lighting, and visibility, and the existence of a slew of different drone types and capabilities render pure simulation-based validation inefficient and often infeasible. 
 
To alleviate this issue, simulations are complemented by field tests, which execute and observe several scenarios, such as drone missions~\cite{bozhinoski2015flyaq} to validate simulation results in the real world.
However, setting up and executing field tests is a rather time- and resource-intensive endeavor that requires preparing hardware, deriving test scenarios, and setting up equipment prior to field test execution.  
In particular, the volatile environment of mission execution contexts, e.g., weather conditions, emergent situations, and changes due to environmental disturbances, makes field testing challenging, requiring approaches for 
(i) structured and systematic execution \textit{guidance}, and  
(ii) \textit{documentation} of adaptations, deviations, and emerging problems.
Furthermore, field tests need thorough planning and validation, as they pose a significant risk to the humans involved. 
For example, during field tests, drones need to maintain safe distances between each other, human operators, and obstacles.
Test scenarios that, for instance, violate safety distances can lead to crashes, loss of hardware, or injuries.

However, tests are typically performed on a rather ad-hoc basis, lacking proper guidance for involved testers and requiring a significant amounts of internal knowledge from testers, drone pilots, and operators.
 Furthermore, documentation of testing steps, as well as  their successful or failed execution is also often missing or insufficient.
Traditional tool support typically includes manually created documents and checklists that define the test scenarios.
However, these are generally limited, as testers have to interpret the feasibility of scenarios in volatile test conditions, possibly adapting scenarios with insufficient documentation.
For improvement, issues observed in the field need to be thoroughly reported and fed back into the quality assurance and engineering processes for testers, developers, and quality assurance managers.
They use the collected data and observed information to track down the root cause of a problem and apply fixes for risk reduction in the field test, the simulations, and systems engineering.

From these goals and limitations, we derive the following research question. \textbf{RQ.} \textit{\rqtext}
We elicit requirements for the knowledge representation of a field test of drone applications from traditional test goals and examples as the basis for configuring an expert information system to guide field testers in conducting test tasks and collecting validated data for test analysis.

To improve the current state of field testing operations, this paper introduces the approach \textit{Field Testing Scenario Management (\acro)} for field testing guidance that is adaptive and 
provides sufficient guidance for field testers as a foundation for efficient data collection.
Our contribution goes beyond existing approaches, as it leverages features of an information system to define and execute test cases and facilitates quality assurance and iterative improvement of field tests and CPSs.

\acro shall leverage concepts from scenario-based requirements engineering~\cite{sutcliffe2003scenario} and Behavior-Driven Development~\cite{smart2023bdd} (BDD) to define \emph{structured and reusable test scenarios}, with dedicated tasks and responsibilities for role-specific guidance.
\acro consists of three major parts to support test engineers, field testers, and the quality assurance manager -- 
(A) \textit{Test Design and Management:} 
An information system shall facilitate defining and managing structured, flexible, and reusable test scenarios.
(B) \textit{Field Test Execution:} 
A field test application shall execute these test scenarios to provide role-specific guidance and to collect field data.
(C) \textit{Test Analysis and QA:} 
Test data analysis shall process, aggregate, and visualize the field test data in an analysis component.
The three parts shall work together, facilitating end-to-end field test definition, execution, and analysis, with feedback into CPS quality assurance and engineering, e.g., via an issue tracking system. 
We evaluate \acro (with a focus on part A -- the creation of different test scenarios)
in a feasibility study with data from three test scenarios and interviews with three domain experts.

The remainder of this paper is structured as follows.
\citesec{example} presents background and a motivating example.
\citesec{approach} introduces the \acro framework and its usage for creating structured field test scenarios. In \citesec{eval}, we then report on a feasibility study, validating \acro through three application scenarios and expert interviews. Sections~\ref{sec:relwork} and~\ref{sec:conclusion} conclude the paper with related work and future work.

%% file: sec_02_example.tex
\section{Background \& Example}
\label{sec:example}
Thoroughly testing CPSs is a resource-intensive and cumbersome task. It commonly requires carefully executing a significant number of tests in a simulation environment, ensuring that the different parts of the CPS behave according to their predefined requirements. 
For example, for autonomous vehicles, \textit{CARLA}~\cite{carla} provides a powerful open-source simulation environment, allowing complex simulations in different environments with diverse weather conditions and pedestrians. In the domain of drone applications, various simulators facilitate the planning and execution of multi-drone missions, exploring weather conditions, and running high-fidelity 3D simulations~\cite{airsim,gazebo,ardupilot}.
However, a  ``simulator-to-reality'' gap has long been recognized as a significant impediment in CPS and robotics research, describing the mismatch and inaccuracies of simulated and real physical environments~\cite{jakobi1995noise,salvato2021crossing}.
Therefore, it is crucial that, once basic functionality has been ensured, testing needs to transition into the real world to confirm that the system and its constituent components work in the ``real'' physical hardware and realistic environments. 

Field testing is a crucial step before deploying a CPS, such as a drone application, in a real search-and-rescue mission, where a system failure or even a crash of a drone could have severe consequences and even put lives at stake.
While quality assurance by systematically creating and executing test cases for ``traditional'' systems is a well-established area ranging from approaches for fuzz testing, metamorphic testing, or regression, field testing, has been largely neglected.
Only a few tools, such as spreadsheets and shared documents, support this type of activity and even fewer have been designed with a thorough software-quality assurance process in mind~\cite{agrawalRE24}.

As part of our work in the space of drone applications~\cite{cleland2018dronology,vierhauser2021hazard}, we have executed a large number of field tests in different conditions, using various hardware, flight controllers, and software components. Running such field validations typically requires four to six (depending on the number of drones involved) human participants tasked with various activities. 
Each human action is assigned to a specific role such as a \textit{remote pilot in charge} (RPIC) serving as a backup pilot, in case something unforeseen happens during a flight, a \textit{mission commander} (MC) in charge of running a mission, and a \textit{safety inspector} (SI) ensuring that the takeoff/flight area is clear of humans, and that drones are correctly prepared and activated.  

Preparing and executing tests requires preparing drone hardware, setting flight controller properties (such as maximum altitude, geofences, or remote handheld frequencies), creating test scripts, and executing them in the field. %%%%
Tests typically cover both, ``simple'' standard functionality, such as takeoff and hovering, ensuring the system's basic flightworthiness, and complex scenarios, such as synchronized multi-drone takeoff, navigating several waypoints, or returning ``home'' coordinated.
It is important to note that executing field tests in the real world, even when thoroughly planned, always exhibits a certain degree of variation and flexibility to be considered.  
Environmental conditions, such as wind,  are not 100\% reproducible, and testers need some flexibility to account for these factors. 
% Based on our own experience, most field tests rely on a checklist-style approach where key testing aspects are specified, e.g., in a spreadsheet, basic results are captured verbosely as text, and observed issues are scribbled down in various documents. 
% This, however, poses the threat of low data quality, data gone missing, and increased aggregation and data integration effort once a test has been conducted. 
% In order to elevate field testing, and transition from a rather ad-hoc approach to a structured testing process, requires improved method and tool support that considers the specific characteristics of field testing.
From our own experience in field testing and workshops with domain experts, 
we define the following four core capabilities (C1-C4) that we deem crucial for such a structured field testing process.
%\vspace{1em}

(C1) \textbf{Formal, yet flexible, test description.} 
\emph{Tests shall be defined in an actionable, reusable, and concise manner that captures relevant pre-conditions, tester actions, and expected observable outcomes}.
This allows 
% transitioning and 
validating initial test scripts before going to the field to a test drive, with reusable and clearly defined tests that can be bundled in pre-validated test suites for field execution. 
However, tests also require a certain degree of flexibility, so that steps can be defined on a more detailed or abstract level depending on the type of activity and data collection requirements.

(C2) \textbf{Test guidance for diverse actors and roles.} 
\emph{Each test and constituent step requires roles associated with them}.
For example, a mission commander, pilot, or test observer shall perform different actions during a field test, e.g., start a mission, prepare the drone, or observe its behavior. 
This ensures, during field tests, that there are clearly defined execution responsibilities, and that participants are aware of their mission duties, tasks, and events.
This requires each role to receive a list of relevant next tasks, 
which consider the context of mission, role, state of the CPS, and environment. 
Furthermore, a test activity shall be connected to the \emph{relevant resources}, e.g., which button to press, or which switch on a handheld to flip, 
to provide visual cues and avoid overwhelming humans with many complicated task details, in addition to the often mentally challenging field experiments.

(C3) \textbf{Test documentation and feedback.}
\emph{Testers shall get tool support to report deviations from expected behavior} (either for a specific step, or related to a certain event), including issues that prevent conducting the field test as planned, e.g., shortcomings in the test case, data, systems, or coordination. 
After finishing the field test, \emph{analysis support} shall provide the data analyst, developer, or quality engineer with information on which parts of a comprehensive test have been completed successfully, and where issues have been reported.

(C4) \textbf{Validated field test data collection.}
\emph{Test execution information shall be 
(i) structured according to the field test tasks conducted and 
(ii) 
% related 
validated with diverse other sources}, such as flight controller logs or system outputs, as a foundation for traceable test data analysis, in particular considering adaptations of field test scenarios to changing conditions in the test environment. 

%%%%%%%%%%%%%%%%%%%%
%%adaptation/guidance/documentation...
% Executing 

% \sbi{emphasize capabilities for field test adaptation, guidance, documentation to support test result analysis and improvement on levels: test result, test data analysis, test improvement, CPS engineering and application improvement; CPS ease of use, risk reduction}
% \mv{how field testing works in practices.. what is missing - what are core requirements to a new system?}

% RELWORK/Application Areas:

% Zhao, X., Wang, W., Wen, L., Chen, Z., Wu, S., Zhou, K., ... Wu, C. (2023). Digital twins in smart farming: An autoware-based simulator for autonomous agricultural vehicles. International Journal of Agricultural and Biological Engineering, 16(4), 184-189.

%% file: sec_03_approach.tex
\section{The \acro Framework}
\label{sec:approach}

Based on the aforementioned capabilities, we have developed an initial version of the \acro framework.
\acro consists of three main parts (cf.~\citefig{fw}) intended to guide test managers, testers, and QA engineers through the process of managing, executing, and analyzing field tests.
Part (A) focuses on the structured and efficient design and management of test scenarios; 
part (B) concerns providing support for test execution in the field; and 
part (C) is dedicated to providing support for consolidating data and analysis.
For the purpose of this paper, we mainly focus on part (A), the design and management of field tests, specific characteristics, needs, and requirements of field tests, particularly for drone applications.

\begin{figure}[t!]
  \includegraphics[width=1.00\linewidth]{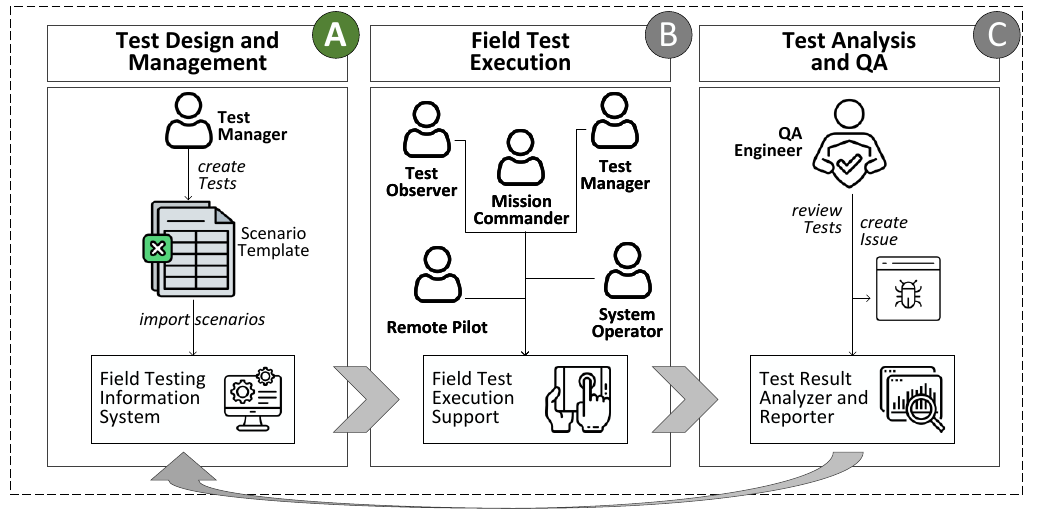}
  \caption{\acro Framework overview.}
  \label{fig:fw}
  \vspace{-1.21em}
\end{figure}

\subsection{Test Management}
Executing tests with drones in the field requires thorough planning and preparation. 
As with any CPS, both hardware and software need to be provisioned, and testers need to ensure that relevant parts (e.g., different types of cameras for testing computer vision components) are available in the field. 
The core part of \acro consists of a test management information system that shall facilitate creating and maintaining test scenarios. 

When performing drone field tests, several actors, such as a mission commander and drone pilots, execute a series of test steps, observe the behavior, and take notes if a test (or a constituent step) is not performed as expected.
We incorporate requirements engineering and testing approaches, combining concepts of Use Case Templates~\cite{yue2013facilitating,robertson2000volere} for more structured test definition, with Behavior-Driven Development~\cite{smart2023bdd} to enhance the test scenarios' expressiveness (cf. C1).
\citetable{scenario} shows an overview of the \acro Field Test Scenario Template.

The first part (\emph{Process Setup}) contains the basic description of the Test Scenario, similar to a Use Case description, defining the name, rationale, and primary/supporting actors. 
These actors are also the roles relevant for performing test steps and tasks (cf. description below).

\textbf{Behavior Description.}
The second part of the scenario is dedicated to describing the steps of the use case and the expected behavior.
BDD~\cite{wynne2017cucumber} is frequently used in agile development to go beyond the description of user stories, specifying testable scenarios in a simple syntax: 
(i) \emph{Given}, the state before the test begins;  
(ii) \emph{When}, the behavior specified as part of the test; and 
(iii) \emph{Then}, the changes expected after the behavior is executed. 
In the context of the \acro field test scenario description, \emph{Given} describes the pre-condition to be met before the test task can commence.
For example, before a pilot can place a drone in the initial launch position, it has to be available at the testing site and ready for use. 
This pre-condition has to be confirmed to initiate the step. 
\emph{When} describes the action to perform, for example, placing the drone on the ground, or scheduling a mission in the mission planning user interface. 
Finally, \emph{Then} specifies the post-condition to be confirmed for a successfully completed field test step. 	

\textbf{Priority.} 
While actors may require guidance throughout the test, following a strict step-by-step approach is not always possible, impeding flexibility required to accommodate for environment variability. 
Some steps need to be executed in a specific order (cf. Given pre-conditions), whereas others may be executed in parallel or in arbitrary order. 
To address this, in \acro, a role shall see and may start all tasks that have their pre-condition fulfilled.
Hence during test execution, using the \acro information system, testers are presented with several tasks that they can start in the current state of the mission. 
The task's priority specified during the test scenario design (which might be adapted at runtime) provides guidance on tasks to consider with high priority while leaving the expert the freedom to execute any of the available tasks.

\textbf{Phase.} 
The phase element specifies which part of a test execution a step belongs to. 
This helps to determine responsibilities and allows further grouping of individual steps.

\textbf{Responsible.} 
This part defines the person/role who is responsible for performing actions in a certain step.
For example, the {\small\texttt{mission commander}} is responsible for starting a mission in the user interface, while a {\small\texttt{drone pilot}} might be responsible for performing an emergency landing in case a problem occurs during the mission (cf. C2).

\textbf{Duration.} 
The duration element allows to constrain a step's duration and guides all participants about the expected duration of individual steps.
For example, for a step where a drone has to reach a certain waypoint, a duration of {\small\texttt{x}} minutes may be specified. 
If this duration is exceeded the responsible role can be notified to confirm that the step has been executed successfully or to further investigate if the drone, e.g., has not yet reached its waypoint.
This allows for detecting potential test failures early, and timely reacting to problems. It is important to note, however, that in the context of these tests, real-time behavior is not a primary concern, and duration is specified in seconds or minutes (rather than, for example, milliseconds).

\textbf{Sub-process.} 
Finally, a sub-process allows the creation of reusable building blocks where certain functionalities can be grouped.
For example, arming an sUAS always consists of three steps, where the pilot activates the remote handheld, presses a button, and visually confirms the arming status.
Specifying sub-processes can help reduce the size and clutter of test scenarios, and improve test maintainability and reusability.

\textbf{Step Multiplicity.}
A key feature of the \acro behavior description is specifying variables and multiplicities (marked with ``{\small\texttt{<\dots>}}'').
This is particularly important if tests comprise multiple drones that perform tasks, with dedicated pilots responsible for a single drone.
For example, in the scenario with three drones (sUAS) performing a synchronized take-off (cf.~\citefig{drones}), the remote pilot assigned needs to perform activation steps for each sUAS: ``{\small\texttt{RPIC(sUAS<x>→<pilot>) shall activate and arm sUAS<x>}}''
This case shall be represented in a single step for each sUAS ({\small\texttt{sUAS<x>}}) and the respective pilot ({\small\texttt{→<pilot>}}).

The concrete actor mappings (sUAS and pilots) are then established when the test is executed (cf. Test Execution).

\textbf{Step Type.} 
To ensure structured test execution with a collection of essential test data, \acro introduces an additional type of steps that can be performed.
\textit{Test data collection} (TD) shall complement the ``common'' step type for \textit{test execution} (TE) that specifies actions directly related to the test scenario. 
% The ``common'' step type for \textit{test execution} (TE) that specifies actions directly related to the test scenario, is complemented by a second type for \textit{test data collection} (TD). 
While TE concerns performing manual actions and operating the user interface or remote handheld, TD shall ensure collecting vital test data at the right point in time, according to the mission state.
For example, a TD step could validate that a task started at a recorded point in time or confirm the number of satellite fixes a drone has before the mission starts (cf. C3 and C4), as a foundation for effective test data analysis. 

\input{table_test}

\subsection{Test Execution}
Once test scenarios are fully specified with respective steps and responsibilities, they can be executed in the field, and individual steps can be assigned to the particular participants.
For this, the \acro information system shall provide the participants each with a dedicated view that contains only the steps relevant to their tasks to be performed, given a run-time mission state.
This shall ensure that field testers can focus on their roles and retain situational awareness, particularly when actions are executed in parallel, to avoid information overload, also referred to as situational awareness demons~\cite{endsley2012}.

\subsection{Test Analysis}
Combining standard test execution steps (TE) and test data collection steps (TD) shall address a major issue of executing field tests without appropriate tool support (cf.~C3).
Traditionally, vital information about issues that occurred during test execution in the field was, at best, partially collected. 
This impeded the analysis of field test data, to precisely relate certain issues to individual test steps, let alone relating issues to precise timestamps in system logs and flight controller data.

\acro shall provide the ability to query the user for test data at specific points in time and allow issue reporting during each test step, as a foundation for collecting and validating test issues and result data associated with test events, e.g., test tasks that occurred at specific points in time.
This shall facilitate in-depth error analysis (cf.~C4) and linking bug descriptions to detailed test data and log file information.

\subsection{Creation of Field Test Scenarios}
This paper focuses on the first part of the \acro framework, the specification of test cases, and their management before executing field tests.
The following provides a brief overview of creating and maintaining test scenarios in the \acro information system. For a drone mission's scope (e.g. performing a  search-and-rescue operation), the test manager, a domain expert, shall describe test scenarios for successful mission outcomes, e.g., locating subjects in a target area, and important mission risks to identify data required to collect, e.g., drone energy state and consumption rate
~\cite{biffl2024configuring,biffl2022risk}.
%~\cite{biffl2022risk,biffl2024configuring}.
The test manager shall efficiently validate the scenario pre- and post-conditions with domain model assets and dependencies~\cite{biffl2021industry,biffl2023validating}. 

Scenarios are then described using BDD language, e.g., the pilot assigned to an sUAS shall arm it.%, or a mission shall be executed by the mission commander.
In an environment that combines human and machine tasks, human tasks translate to \acro tasks that humans can start and finish in the information system;
machine tasks translate to control tasks that may vary from human to fully automated control over a machine.
%
% Therefore, a traditional human task may be conducted in the test by a human or by a machine function, e.g., a cognitive, scheduling, or control function, such as collecting data, choosing a next step, or observing a drone until it has reached a way point, 
% as a foundation for comparing human and machine behavior in the test environment.

The test manager may edit the test scenario in an editor that supports the domain-specific language.
The \acro compiler configures and generates the \acro information system for field test execution, quite similar to Cucumber for JUnit tests but for human tasks.
% \sbi{@KM: comment how \acro is similar to/different from the traditional use of Gherkin/Cucumber}
%
During field test execution, the \acro information system shall represent the mission state, inform the role-specific tester views, and collect data for test analysis.

To maintain test scenarios, the test manager and software engineer shall use their software engineering means for version/configuration management of a test suite.

%%%%%%%%%%%%%%%
% \mv{this part is still a bit short...}
% % traditional: normal-course requirements
% % test case + rucm + bdd + marker, 
% % => reduce cognitive load
% % => pre-conditions (Tasks+ runtime conditions)
% % assertions -> assertion-based programming...
% % linked to code.(trace links to code/process blocks... - interface conditions)=> identify relevant parts for testing
% % (all different types of artifacts)

% % sample test specification table
% % implicit knowledge is explicit in test description (observations..)

%% file: table_test.tex
\begin{table*}[]
\footnotesize
\caption{\acro Field Test Scenario Template}
\label{tab:scenario}
\addtolength{\tabcolsep}{-0.4em}
\begin{tabular}
{|p{0.5cm}|p{2.3cm}|p{4.5cm}|p{3cm}||p{1cm}|l|l||p{1cm}|p{2.2cm}|}
\toprule
\multicolumn{8}{c}{\textbf{Process Setup}} \\
\midrule
\multicolumn{2}{l}{\textbf{ID}}   & \multicolumn{7}{l}{Test-Case 01 (TC01)}\\
\multicolumn{2}{l}{\textbf{Name}}  & \multicolumn{7}{l}{Test Case-02: Multi-sUAS Synchronized Takeoff}    \\
\multicolumn{2}{l}{\textbf{Description}}  & \multicolumn{7}{p{14.5cm}}{Multiple sUAS shall perform a synchronized takeoff check, flying to a waypoint and RTL (all RTL altitudes must be different).} \\
\multicolumn{2}{l}{\textbf{Primary Actor}}                         & \multicolumn{7}{l}{mission\_commander, sUAS} \\
\multicolumn{2}{l}{\textbf{Supporting Actor}}                    & \multicolumn{7}{l}{pilot\_1, pilot\_2, pilot\_3}  \\
%\multicolumn{2}{l}{\textbf{Parameters}}     & \multicolumn{7}{l}{} \\
\midrule
\multicolumn{8}{c}{\textbf{Process Definition}}\\\midrule
\emph{Step} & \emph{Given}  & \emph{When}   & \emph{Then} & \emph{Type} & \emph{Resp.}  & \emph{Prio.}  & \emph{Sub.Pr.}      & \emph{SubPr. Params} \\\midrule
1x.1                                             & sUAS\mar{x} is available at test site. & RPIC(sUAS\mar{x}$\rightarrow$\mar{pilot}) shall place sUAS\mar{x} in its launch location. & sUAS\mar{x} is placed in its launch position. & PC (M) & \mar{pilot}    &          &                &                                                             \\
1x.2                                             & sUAS\mar{x} is disabled.& RPIC(sUAS\mar{x}$\rightarrow$\mar{pilot}) shall activate and arm sUAS\mar{x}.            & sUAS\mar{x} is activated and armed.           & PC (M) & \mar{pilot}    &          & sUASArm & \{"sUAS": "sUAS\_x", "pilot": "sUAS\_x \textgreater pilot"\}\\
\bottomrule
\end{tabular}
\vspace{-10pt}
\end{table*}

%% file: sec_05_eval.tex
\section{Validation - Scenario-based Feasibility Study}
% \section{Evaluation - Application Example}
\label{sec:eval}
To answer our research question \emph{``\rqtext''}, we (i) assess the general feasibility of our framework including the suitability of \acro to support field testing, and (ii) collect additional insights and evidence into how testers perceive our approach, and collect further requirements that have not been initially considered.

For the first part, we use examples from a multi-sUAS system that we have been developing for several years and deployed in the real world~\cite{cleland2018dronology}, collecting initial evidence that our approach allows modeling real-world test scenarios, considering all necessary steps, conditions, roles, and data collection needs.
For the second part, we conduct interviews with three drone system researchers with extensive development and field testing experience, asking them about the potential suitability of \acro for field testing.

\subsection{Case Project: A Multi-sUAS Application}
In the past several years, as part of our research, we have been engaged in the development of several versions of a system for drone mission planning and execution~\cite{cleland2018dronology,agrawal2023requirements}. 
The system is fully deployable using both physical and simulated drones and has been used for several case studies and demonstration applications, such as river search-and-rescue, aerial photography, and emergency response missions~\cite{cleland2018dronology,agrawalRE24,cleland2020requirements}. The system comprises various user interfaces, components for mission planning and execution, a \textit{Ground Control Station} (GCS), and onboard computation capabilities on the individual sUAS for performing tasks (semi-)autonomously~\cite{chamberssesos24}.
Throughout the lifetime of the system, we have performed extensive testing, both with simulated drones using high-fidelity simulation environments, such as \textit{Gazebo}~\cite{gazebo} or the \textit{Ardupilit} Software-In-The-Loop (SITL) simulator~\cite{sitl}, and real hardware at flying fields (cf.~\citefig{drones}).
Every time new functionality is developed, or a new application use case is introduced, a step-by-step process of thorough software testing, simulation runs, and field tests is required to confirm that the new features do, in fact, work in real-world settings.
This, in turn, requires thorough preparation of field tests, ensuring that the hardware, alongside the software, is available at the flying field, and that a series of test scenarios is executed and documented in a sufficient level of detail.

\subsection{Evaluation Setup}
% % In this section,
% % In the following section, we describe we describe the 
% This section describes three application scenarios 
% \sbi{re-check with final section content}
% to validate the first part of the \acro framework, the creation and management of test scenarios, and the interviews we conducted with sUAS application developers and testers.

% \bullitem{I -- Scenario-based Validation:}
To assess (i)  the general feasibility and expressiveness of \acro, and the extent to which detailed tasks and conditions can be depicted with the proposed concepts, we selected three application scenarios, where our drone system was used in the past, and where we have conducted extensive field tests~\cite{agrawal2023requirements,ChambersHiFuzz2024}. One researcher familiar with the drone system, who has also conducted several tests in the field, created detailed test scenarios for each use case, and two other researchers subsequently validated the test scenarios and provided feedback, comments, and suggestions for improvements.
The goal of this first part was to assess whether the proposed concepts are sufficient, or important aspects of a field test that cannot be modeled and described with \acro.\footnote{We provide scenarios and evaluation materials as supplemental material in our GitHub repository:~\url{https://github.com/SE-CPS/FiTs_public}}

In a second step, we investigated the (ii) practical applicability of \acro, and whether developers deemed the proposed scenario-based test description useful or whether crucial concepts were missing or overlooked. For the semi-structured interviews, we created a questionnaire with three parts. The first part assesses the current state of practice, asking about current tools and approaches used for planning and executing field tests.
The second part, briefly introduces the \acro framework, the main concepts, and its usage. 

Finally, in the last part, we ask participants whether they could imagine using such a structured test description in practice if the proposed workflow would be suitable, and if any key features or capabilities are missing.
In total, we recruited three participants, each of whom had extensive experience in both developing software for drone systems and performing field tests.

\begin{figure}[t!]
\centering
\begin{subfigure}{0.24\textwidth}
    \includegraphics[width=\textwidth]{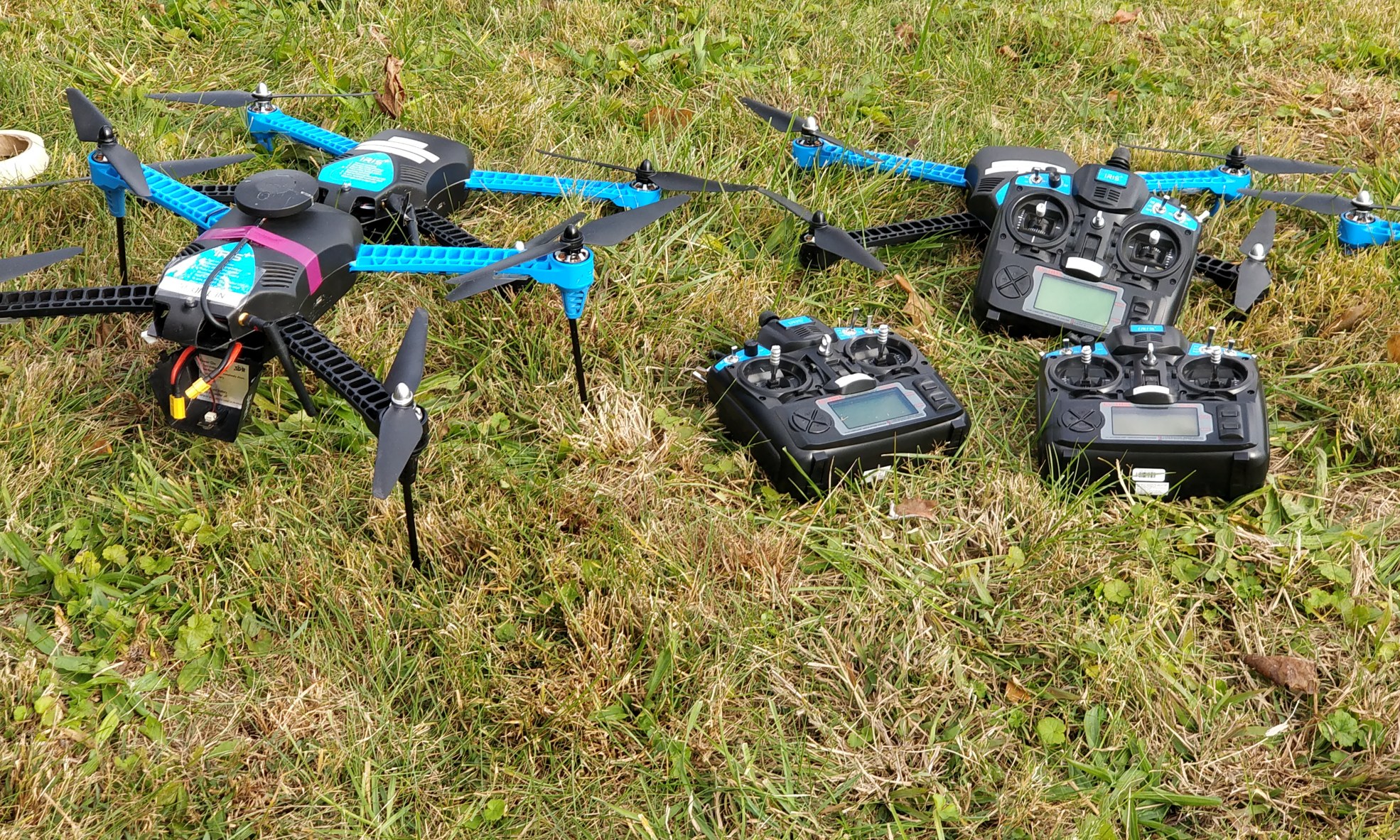}
    \caption{Field Test preparation, before test execution with three drones.}
    \label{fig:first}
\end{subfigure}
\hfill
\begin{subfigure}{0.24\textwidth}
    \includegraphics[width=\textwidth]{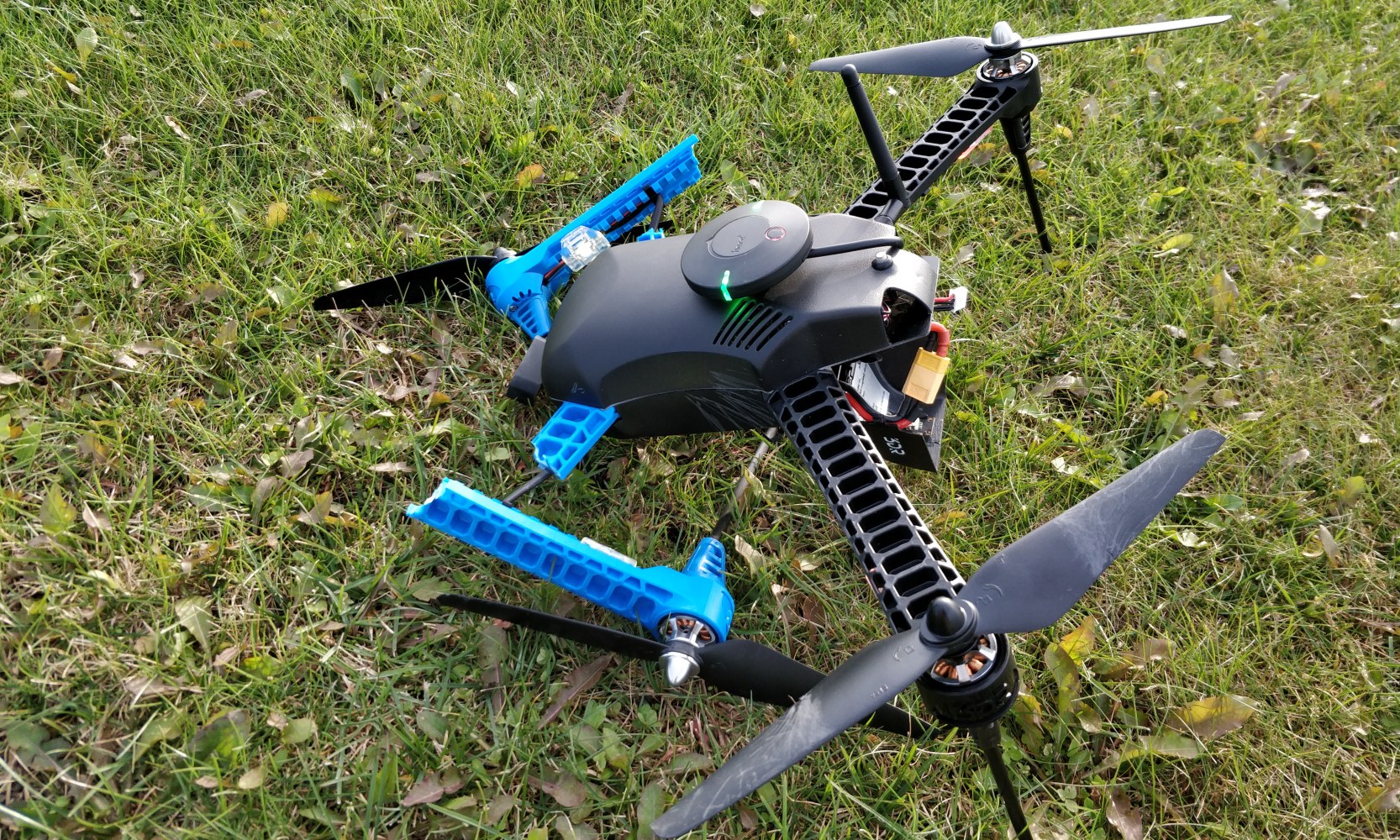}
    \caption{Damaged drone as a result of an unsuccessful Field Test.}
    \label{fig:second}
\end{subfigure}
\hfill   
\vspace{-15pt}
\caption{Multi-drone field test equipment.}
\label{fig:drones}
\vspace{-1.5em}
\end{figure}

\subsection{Results}
 In the following section, we describe the results and further discuss the findings and lessons learned.
 
\textbf{(1) Scenario Creation:} 
As part of the scenario-based validation, we successfully created the necessary tasks, conditions, and roles for all three scenarios. 
The first scenario represented a simple take-off test (21 steps), the second scenario represented a more complex multi-drone scenario, testing synchronized take-off of three drones (36 steps), and the last scenario depicted a more coarse-grained mission execution test (20 steps).
Altogether, it took about 45 to 60 minutes per use case to create the respective field test scenario with all steps, roles, and conditions. 
While for the simple cases, the resulting test tasks and activities are rather concise, for the third, more complex test scenario, the template is already quite extensive, resulting in 92 tasks (including exception states and alternatives).
The \acro template provided the freedom to specify crucial steps in more detail, e.g., regarding requirements during multi-drone takeoff, whereas other steps allow for more flexibility, such as the preparation and inspection of drones prior to take-off, where some steps might be executed in arbitrary order -- not requiring strict step-by-step instruction.
Furthermore, dedicated data collection tasks allowed for collecting data, such as modes of the drone or weather conditions at specific points in time during the test, ensuring that important data was documented. 

The initial results indicate that practitioners can take up \acro to validate field test scenarios and consider the required guidance for testers.
\acro also addresses adaptation needs, due to changes in the field test environment, and that field test documentation meets regulation requirements and supports data analysts, quality assurance experts, and CPS engineers.

\textbf{(2) Expert Interviews:} 
During the interviews, all participants acknowledged shortcomings regarding process and tool support for preparing and executing sUAS field tests in a structured and reproducible manner. 
The participants confirmed that traditional approaches for field test specification mostly rely on spreadsheet-based checklists and shared documents to prepare instructions for field tests and to collect notes during test execution.
Regarding field test result data collection, one participant mentioned that it is a ``\emph{logistical challenge to [...] combine the data from all the different sources, such as log files, and combine them with what was going on during a test, ..., information easily gets lost''} (cf. C4).
Furthermore,  participants emphasized that the field test execution can easily become quite complicated, requiring a significant amount of training and internal knowledge about different steps and activities, such as specific parameters and procedures. 
Furthermore, participants also noted that the structured approach can help to better collect information during the test and provide feedback to developers (cf. C3). For example, if a human could not understand something in the user interface (or was missing certain functionality), this could be directly reported, and related to the specific step in the test. 
One participant also mentioned that such a structured approach (cf. C1) can greatly improve training and preparation for the field tests, particularly for \emph{''participants that have not been part of field testing before''}. 
Altogether, the feedback we received during the interviews was very positive with several suggestions for further improving the scenario description.

\subsection{Discussion}

% Our 
The empirical evaluation has confirmed that 
(1) the \acro structured test scenario description can be successfully used to describe real-world test scenarios with different roles, responsibilities, and tasks, and 
(2) the \acro framework is likely to provide additional value for field testers by facilitating structured and systematic field testing. 
The test scenarios were easy to create and customize through the use of the predefined template spreadsheets.

The interviews further provided insights into additional functionality, particularly to increase the ease of use of the approach. 
One participant mentioned that ``\emph{automation support for selecting and creating scenarios and steps [...] e.g., in the form of a wizard-based approach, would be particularly helpful}'' to reduce the additional effort 
% that is 
required to initially create the task.
However, while the participants mentioned the additional time required for the field test, they also acknowledged that this would pay off 
% in the long run 
once a set of reusable scenarios is available for testing.

Along this line, one aspect that needs to be further validated pertains to the scope and size of the test scenarios. 
One participant mentioned that when creating test cases one needs to be careful to avoid adding too many additional tasks for testers. Executing field tests is already a challenging and labor-intensive undertaking, and extra tasks for data collection/validation might overwhelm testers.
This issue could be addressed by adding an additional validation/consistency checking step before a created test scenario is added to the information system, and deployed in the field.
The test manager can then review the tasks and responsibilities for each role and make sure that testers can focus on their primary responsibilities and are not distracted by too many additional data collection requirements. 
\acro could 
% also 
provide the means for (remote) test observers to take part in the test, who may collect data in the field or review live data in the office, and guide the field testers in case they observe an issue. 

A second avenue for extending the \acro framework was mentioned during the interviews, which is related to the great potential for connecting/integrating diverse artifacts besides issues and log data. 
Particularly establishing traceability to the actual requirements that are validated as part of the field tests could be beneficial. 
This would further foster a thorough safety assurance process%~\cite{clothier2017making}, 
where \acro structured field tests can serve as concrete safety evidence for a safety assurance case~\cite{HawkinsHK13}, in conjunction with other artifacts, such as unit tests and simulation runs that were executed prior to the field tests.

% \textbf{Test Variability Management}
%\bullitem{Test Variability Management}
Variability models~\cite{pohl2005software} can describe the scope of context, in which test scenarios can be applied/reused to efficiently cover a potentially large solution space. 
% \mv{add something about varability and contexts? (do we need it/have the space for that?)}

\subsection{Threats to Validity}
As any empirical study, this 
% Our 
work is subject to 
% a number of different 
threats to validity. While we have demonstrated that the scenario-based structure of \acro can be used to model real-world test scenarios based on our own experience, the scenarios are limited to three application cases. Furthermore, additional external evaluation of the process is required to ensure its applicability in a broader scenario.
However, when asked about the generalizability of the approach, all participants mentioned that any kind of CPS, where field testing is performed, such as autonomous systems, would benefit from an approach like \acro. 
We are, therefore, confident that \acro could be applied to diverse types of systems and scenarios. 

Further limitations may stem from the selection bias of interview participants and their areas of expertise. 
While all participants had extensive knowledge of drones and field testing, they are related to the same application domain. 
As part of our efforts to further validate our work, we plan on conducting an extended user study with more diverse participants including other application areas, such as 
% for example, 
agricultural monitoring or aerial photography.

%% file: sec_06_relwork.tex
\section{Related Work}
\label{sec:relwork}
Testing CPS is a well-established research area, with a significant body of work. 
Techniques and approaches range from assessing the accuracy of onboard AI models~\cite{chandrasekaran2021combinatorial} to applying software testing techniques, such as Fuzzy and Metamorphic testing, to automatically generate new diverse test cases~\cite{wang2021robot}. 
However, most approaches so far have either focused on different testing strategies, or simulation, with a dearth of thorough support for testing in the field.

\textbf{CPS Testing:}
Several authors have investigated CPS testing techniques and strategies. 
For example, Abbaspour~\etal~\cite{abbaspour2015survey} conducted a survey focusing on testing levels, from low-level hardware testing to system and integration testing. 
A challenge they state is the frequent interaction of CPSs with users and requirements pertaining to implementing and testing these systems. 
Similarly, Zhou~\etal~\cite{zhou2018review} conducted a review on CPS testing methods, identifying ten relevant aspects, such as testing paradigms and technologies, and available testbeds. 
While some testbeds do include both hardware and simulation, testing approaches tend to focus on non-manual activities. 
This confirms the need for supporting testers in the field, providing structured means for creating and executing field tests. 

Investigating more specific testing approaches, Kannengiesser~\etal~\cite{kannengiesser2020behaviour} proposed a BDD method for testing Cyber-Physical Production Systems (CPPSs). 
However, their work focuses on transforming BDD/Gherkin scenarios into CPPS models, while this work targets the structured definition of field tests augmented with BDD concepts. 
Similarly, Porres~\etal~\cite{porres2020scenario} proposed a method for scenario-based testing for ships. 
Their work, however, focuses on creating test cases for collision avoidance to be executed in a simulated environment. 
Maciel~\etal~\cite{maciel2022systematic} conducted a systematic mapping study on robotic testing. 
One of their findings, that only one primary study discusses the impact of the adoption of robots on test case generation, confirms the need for structured testing support in domains besides drone applications.

\textbf{UAV Testing:}
Focusing on testing sUAS systems,  Adigun~\etal~\cite{adigun2022metamorphic} employed Metamorphic Testing for testing single and multi-drone operations. However, their work focused on simulation and software in the loop 
% (SITL) 
testing, rather than field test execution. Similarly, Schmidt and Pretschner~\cite{schmidt2022stellauav} focus on scenario-based testing of UAV applications in a simulated environment, and work by 
Yildirim~\etal~\cite{yildirim2018system} created automation support for system-level testing of UAV applications using simulation.
Engebraaten~\etal~\cite{engebraaten2018field} focus on controller field testing for swarm-based drones. However, their work is concerned with actual field test execution, providing insights on setup and preparation, but lacking a proper testing and quality assurance process. 
Our previous work also focused on establishing testing support for UAV applications, particularly in the context of simulation and human interaction~\cite{agrawal2023requirements,ChambersHiFuzz2024}. However, during this previous work, we also observed the need for and lack of a structured process and test description for field testing beyond virtual simulations. 

\textbf{Other Testing and Validation Approaches:}
In previous work on CPPS validation, 
we validated concepts in production test scenarios with technical dependencies in a Production Asset Network (PAN)~\cite{biffl2021industry,biffl2023validating} 
to ensure the availability of data sources for test conduct and analysis.
To identify likely root causes of a production risk effect that may prevent achieving a mission goal~\cite{biffl2024configuring}, and to apply countermeasures~\cite{biffl2022risk}, we applied risk analysis and software slicing principles to CPPS.
In this paper, we build on these contributions to efficiently validate test scenarios for CPSs, such as drones, and the pre-/post-conditions of these tests.

%% file: sec_07_conclusion.tex
\section{Conclusion and Future Work}
\label{sec:conclusion}
In this paper, we have presented the approach \textit{Field Testing Scenario Management (\acro)} for systematically creating field testing scenarios for drone-based applications.
% Our proposed approach \textit{Field Testing Scenario Management (\acro)}, 
\acro provides a flexible framework for creating and managing diverse test scenarios, which are then executed at the field. 
\acro leverages concepts from scenario-based requirements engineering and Behavior-Driven Development, introducing dedicated tasks and responsibilities for role-specific guidance of human experts, novices, and machine actors.
This shall facilitate 
(i) the collection of data in the field 
with sufficient quality and
(ii) quality assurance and iterative improvement of field tests and the tested CPS. 
An evaluation of \acro with three application cases for search-and-rescue drone missions demonstrated its feasibility.
Interviews with experienced drone experts provided promising feedback on its usefulness and resulted in issues to consider in future work. 
For future work, we plan additional empirical studies with \acro prototypes to explore the strengths of an expert information system for guiding field testers as well as challenges regarding adaptation needs due to disturbances in the field, including sensor errors, control loop issues, and approaches for recovery from typical field testing issues.